\documentclass{amsproc}
\usepackage{graphicx}

\theoremstyle{definition}

\theoremstyle{remark}

\numberwithin{equation}{section}

\def\One{\mathbb{I}}
\def\wthree{\;\raisebox{-26mm}{\includegraphics[trim=5mm 10mm 5mm 10mm,clip,width=24mm]{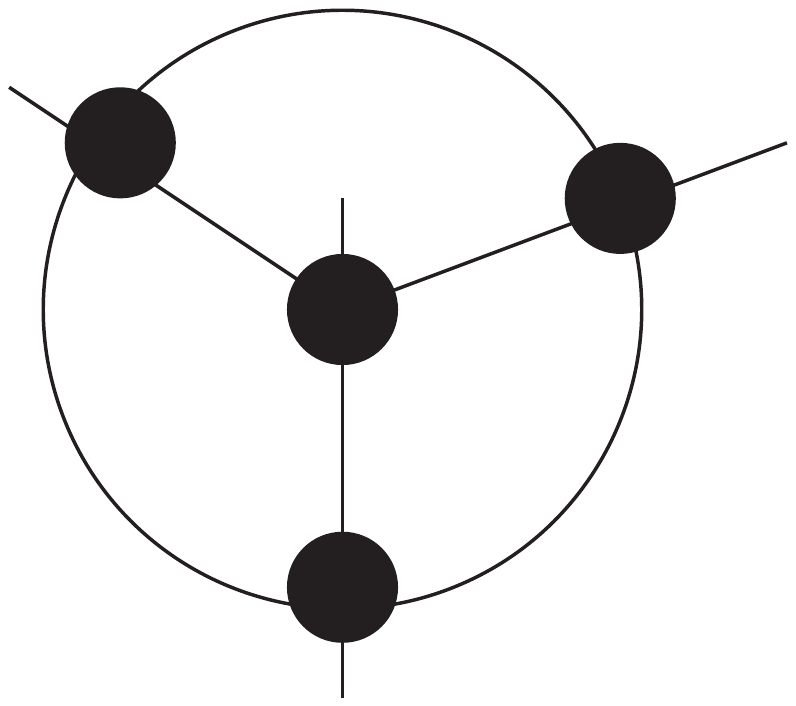}}\;}
\def\wfour{\;\raisebox{-24mm}{\includegraphics[trim=5mm 20mm 5mm 5mm,clip,height=36mm]{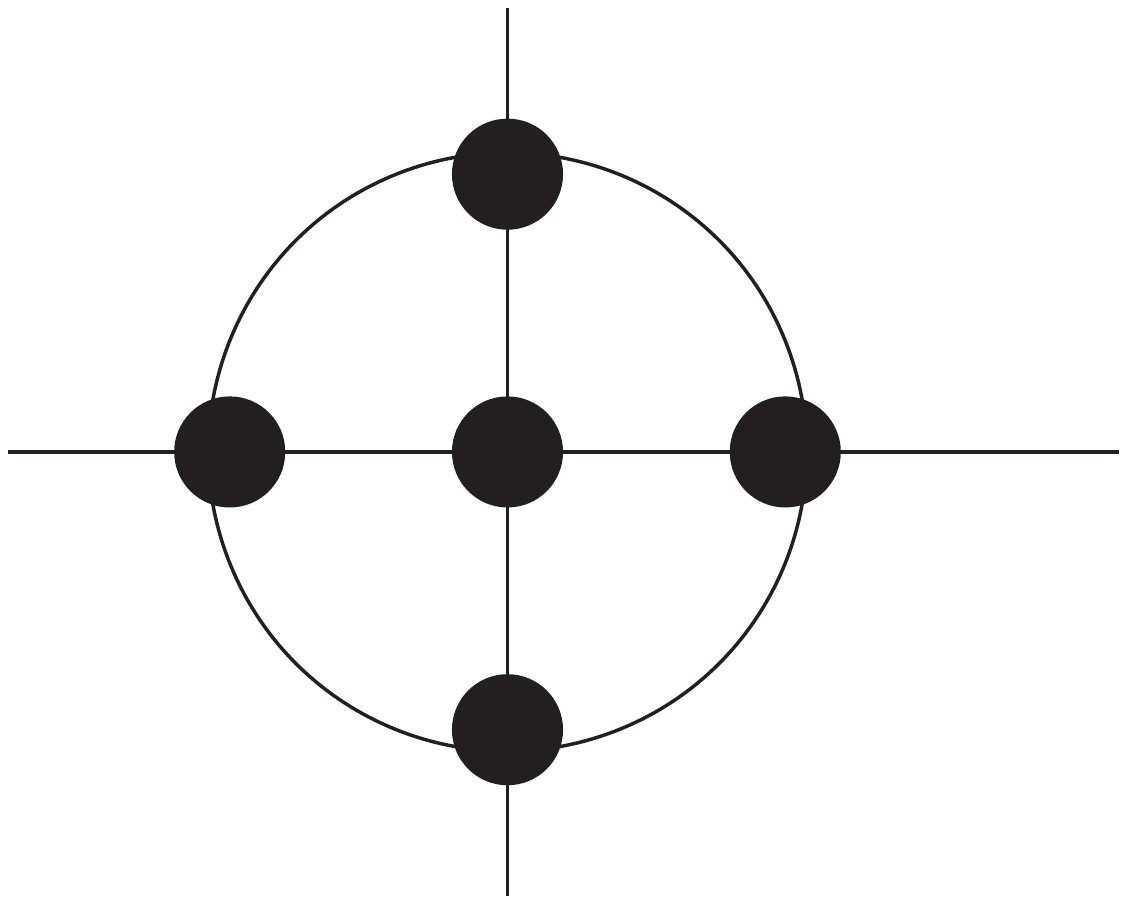}}\;}
\def\wft{\;\raisebox{-24mm}{\includegraphics[height=36mm]{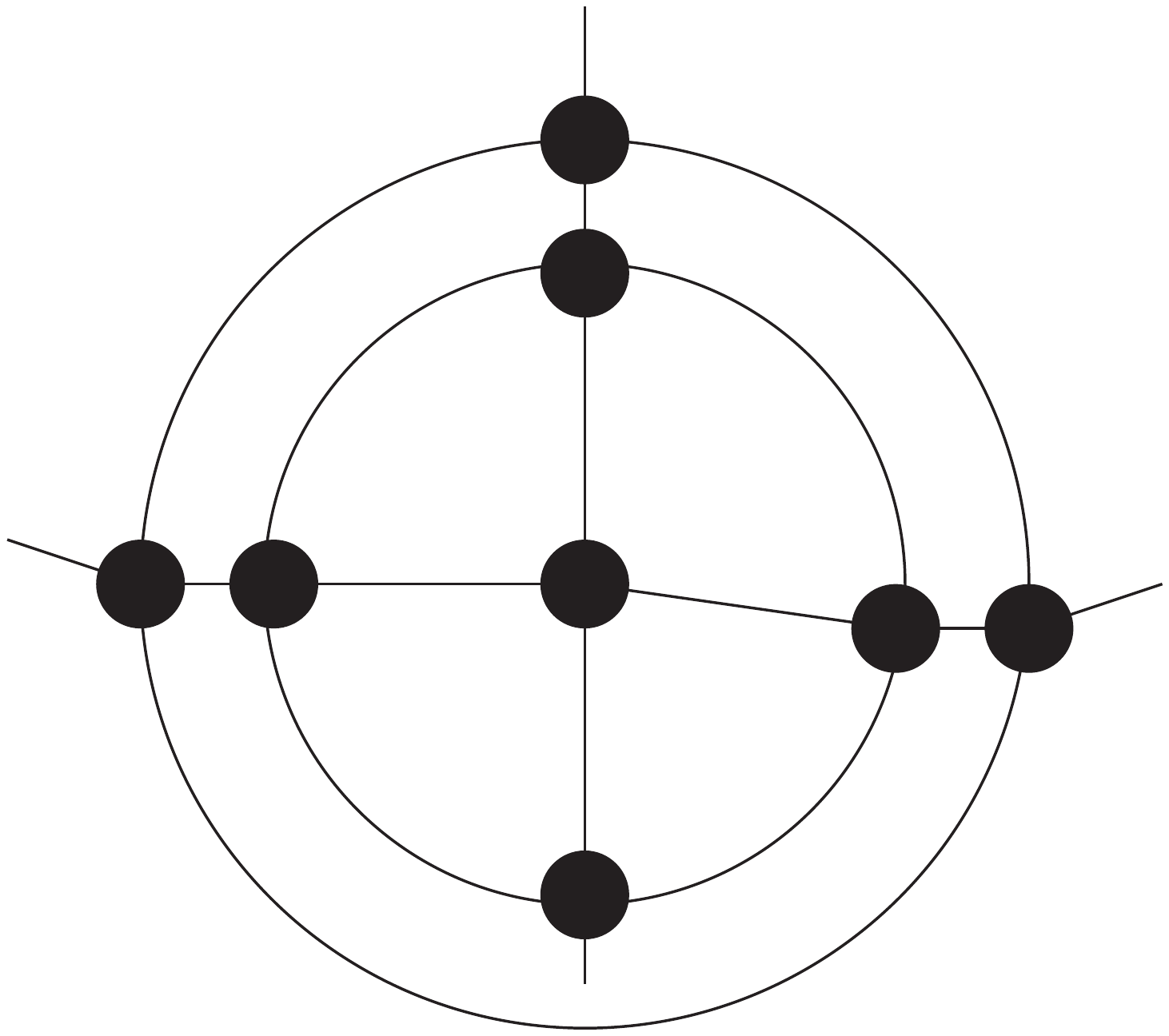}}\;}
\def\wtfa{\;\raisebox{-24mm}{\includegraphics[height=36mm]{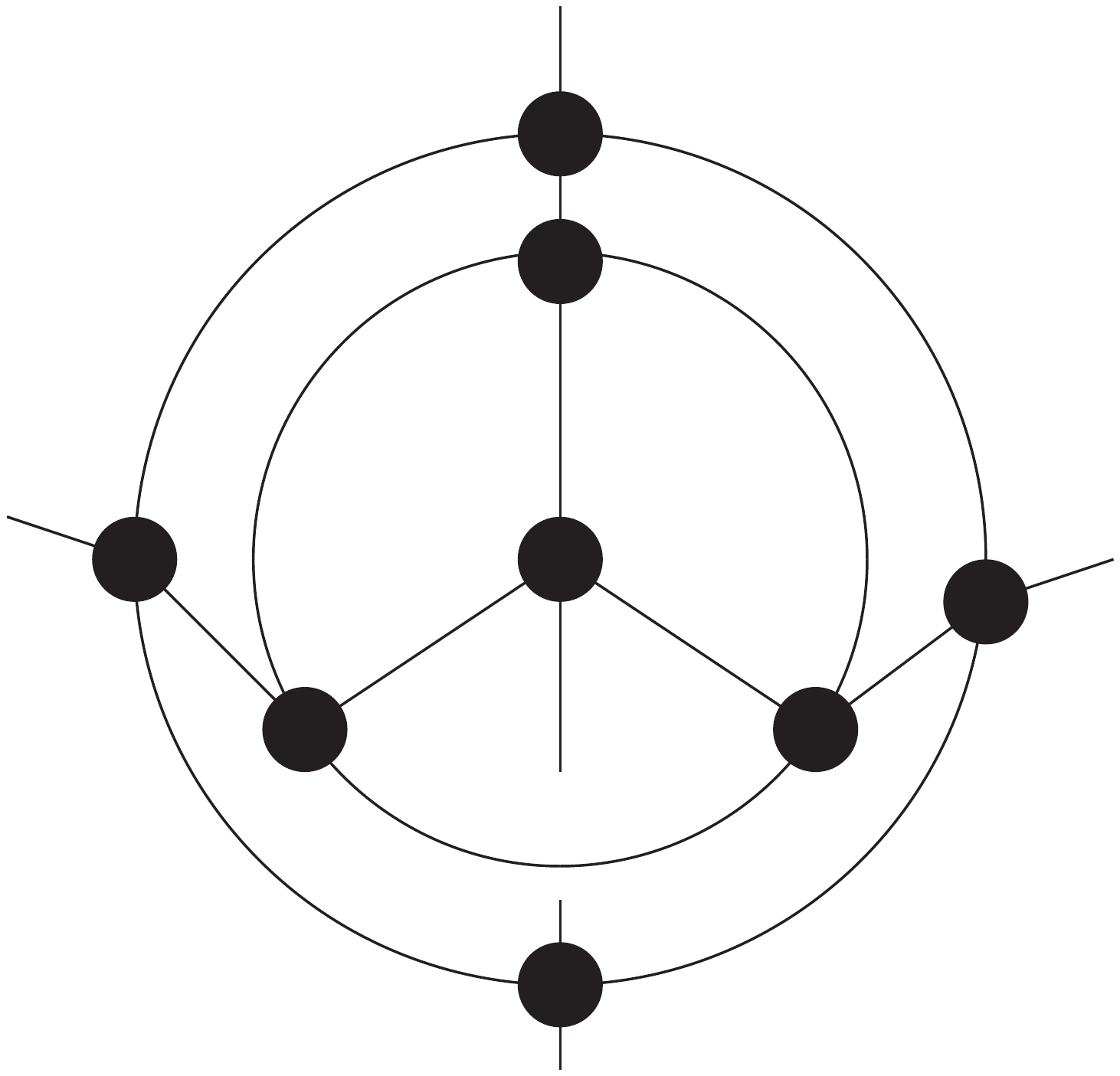}}\;}
\def\wtfb{\;\raisebox{-24mm}{\includegraphics[height=36mm]{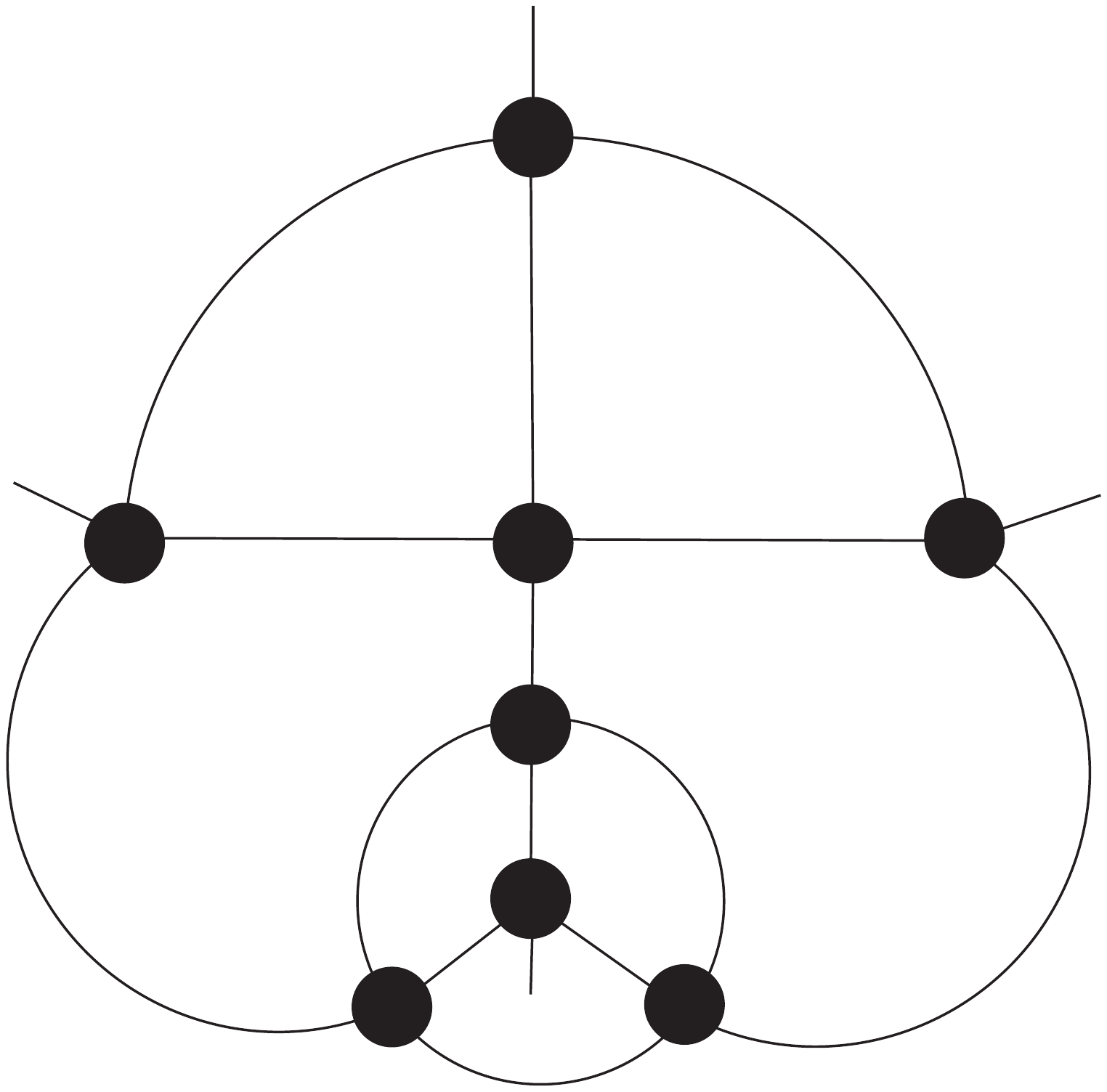}}\;}
\def\wff{\;\raisebox{-24mm}{\includegraphics[height=36mm]{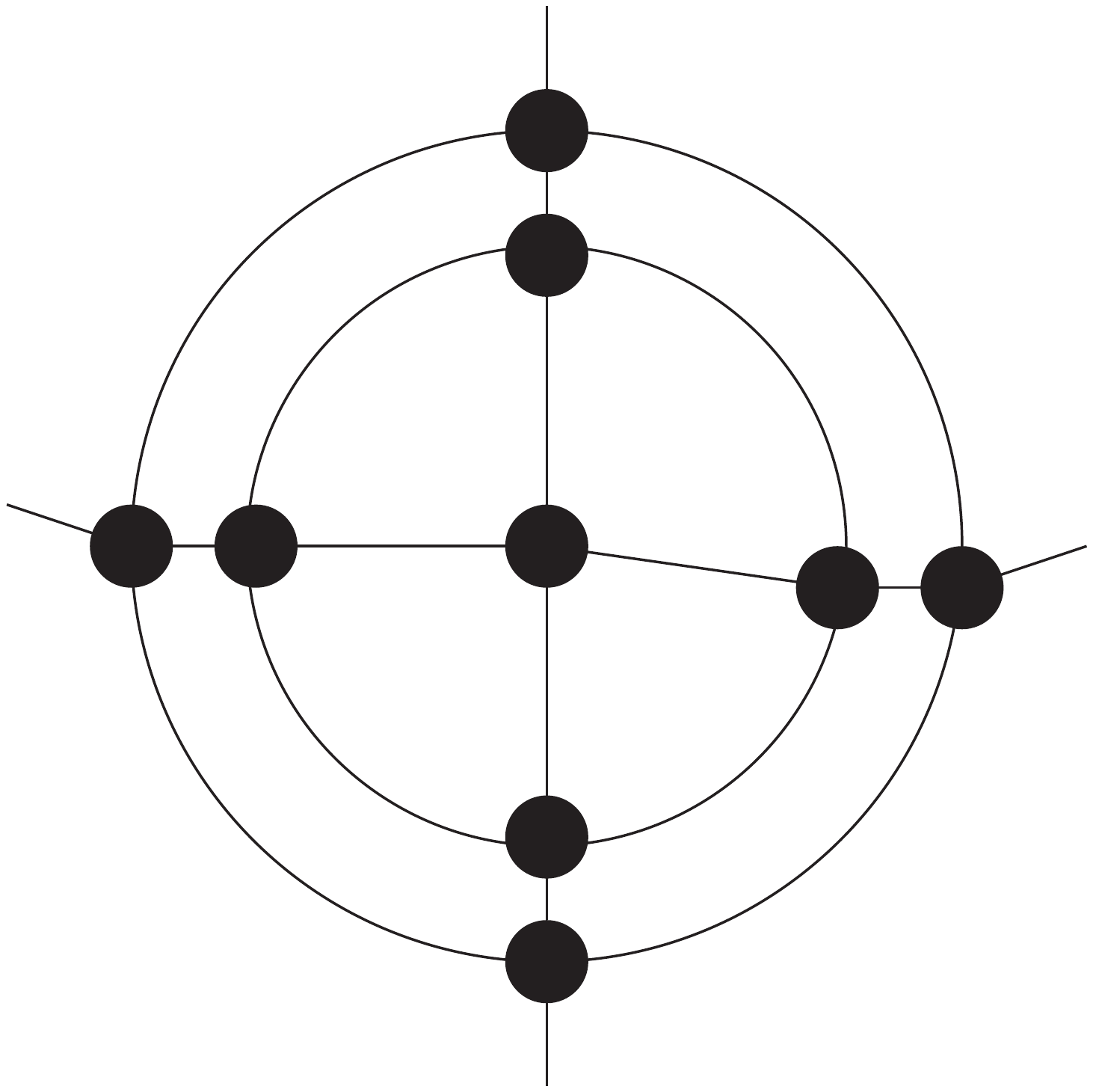}}\;}
\def\wtt{\;\raisebox{-24mm}{\includegraphics[height=36mm]{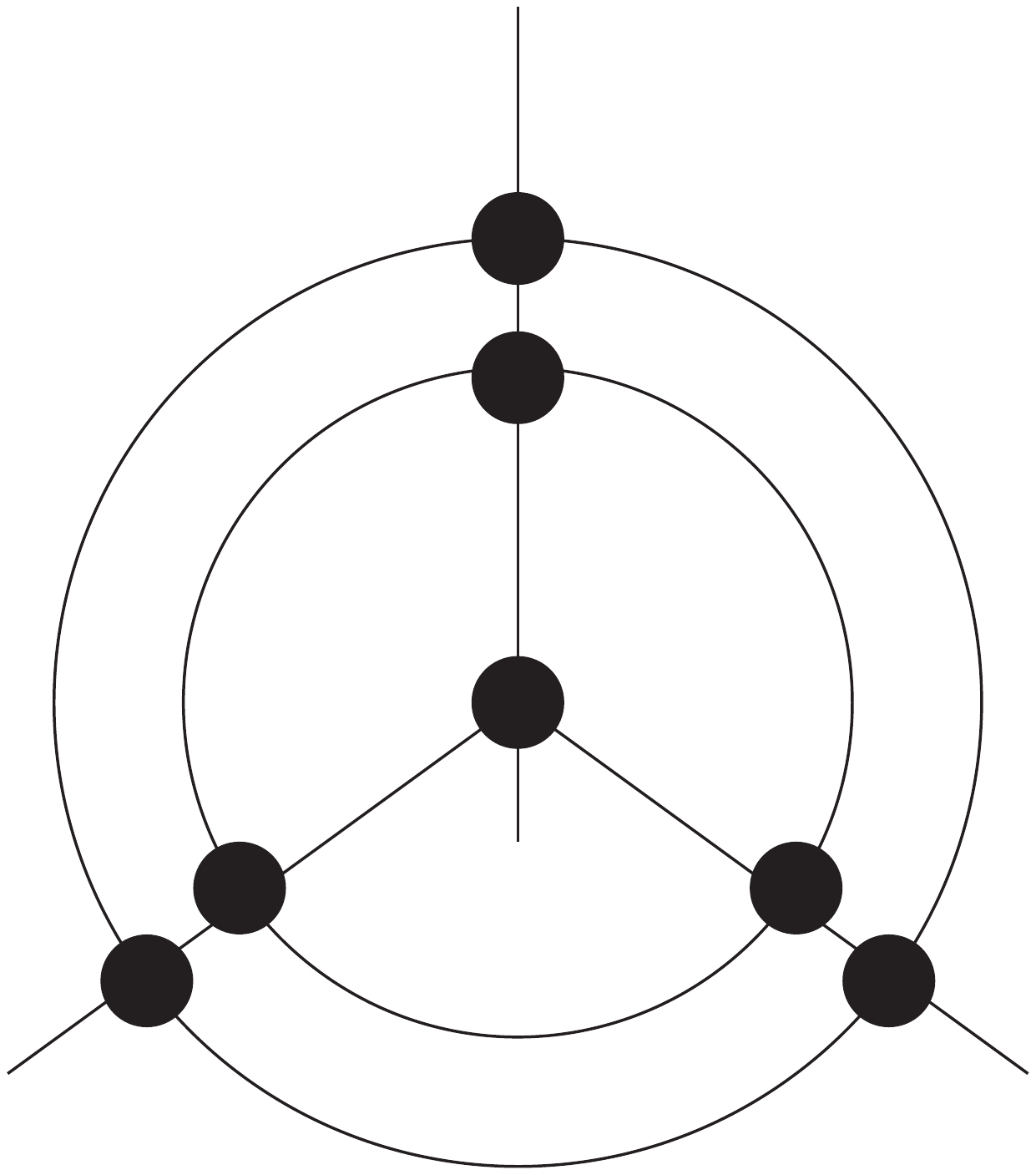}}\;}
\def\gt{\;\raisebox{-24mm}{\includegraphics[height=36mm]{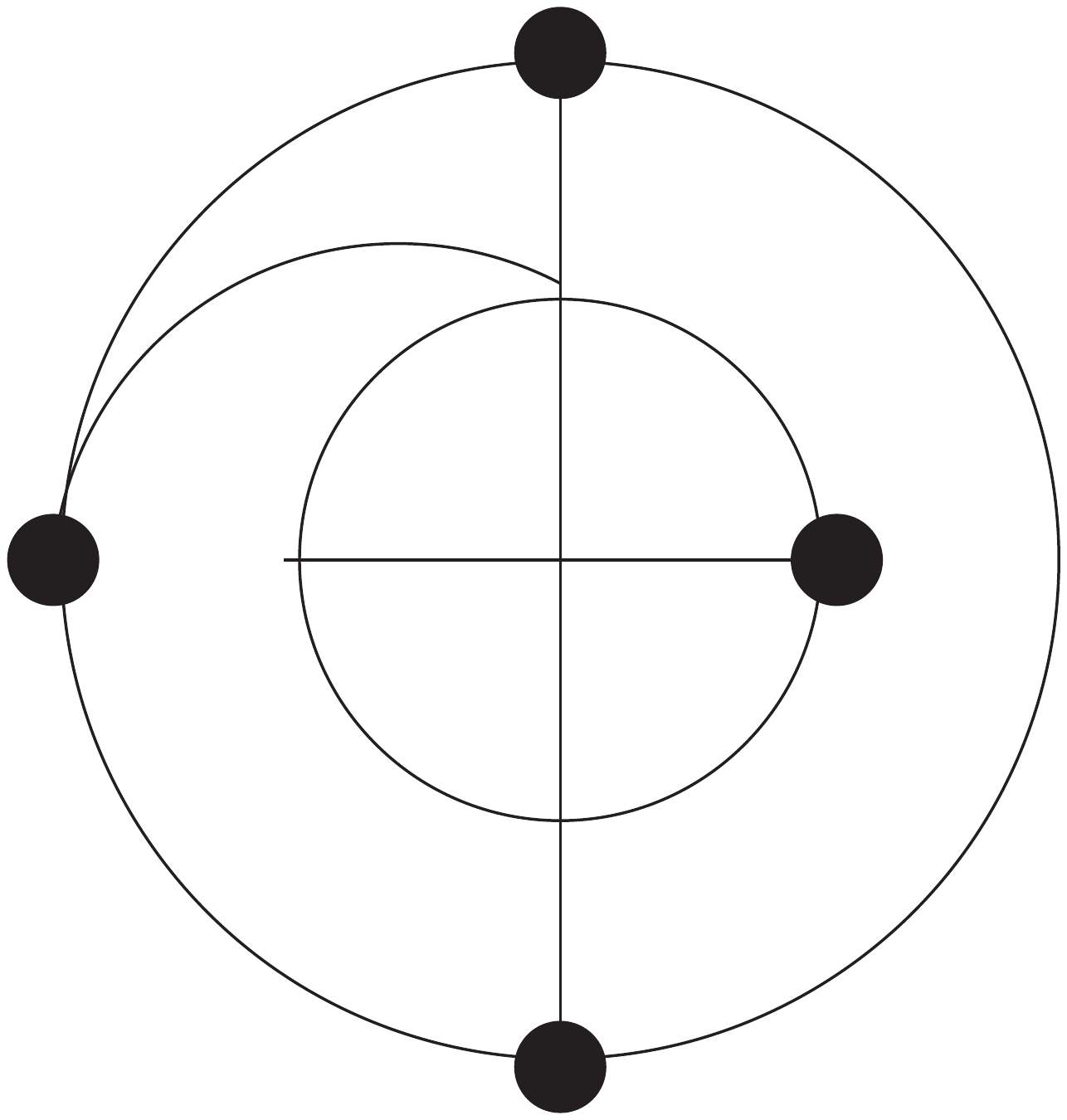}}\;}

\newtheorem{thm}{Theorem}[section]
\newtheorem{lem}[thm]{Lemma}
\newtheorem{prop}[thm]{Proposition}
\theoremstyle{definition}

\newtheorem{rem}[thm]{Remark}

\begin{document}

\title{Quantum fields, periods and algebraic geometry }

\author{Dirk Kreimer}
\address{Dept.\ of Physics and Dept.\ of Mathematics, Humboldt University, Unter den Linden 6, 10099 Berlin, Germany}
%\curraddr{CERN, Geneva}
\email{kreimer@physik.hu-berlin.de}
\thanks{Author supported by the Alexander von Humboldt Foundation and the BMBF through an Alexander von Humboldt Professorship.}

%    General info
\date{March 1, 2013.}

%\dedicatory{_}


\begin{abstract}
We discuss how basic notions of graph theory and associated graph polynomials
define questions for algebraic geometry, with an emphasis given to an analysis of the structure of Feynman rules
as determined by those graph polynomials as well as algebraic structures of graphs. In particular, we discuss the appearance
of renormalization scheme independent periods in quantum field theory. 
\end{abstract}

\maketitle

\section{Introduction}
In this contribution, we want to review work concerning the structure of local renormalizable quantum field theories.
Our emphasis will be to exhibit the most recent developments by way of example, and in particular to stress that at the time of writing we witness two simultaneous developments: a better understanding of the 
algebro-geometric underpinning of field theory in four dimensions of space time, and also as a consequence the emergence of computational approaches which surpass the hitherto established state of the art.
\subsection*{Acknowledgments}
Foremost, I want to thank David Broadhurst, with whom my interest in the periods emerging in QFT started some twenty years ago \cite{Pisa}. Spencer Bloch helped to uncover the mathematics behind it, as did  Francis Brown, Christian Bogner,  Alain Connes, Dzmitry Doryn,  H\'el\`ene  Esnault, Erik Panzer, Oliver Schnetz,   and others. It is a pleasure to thank James Drummond for hospitality at CERN, Geneva, Feb 24-March 01 2013,
as well as the IHES,  where parts of this paper were written.
\section{Graphs and algebras}

\subsection{Wheels in wheels}
It is the purpose of this section to completely analyse an example. 
We choose wheels with three or four spokes, inserted at most once into each other. Results for them are available
by methods which were recently developed \cite{Brown1,Brown2,BrKr1,GraphFun} and which are presented elsewhere \cite{PanzerBingen,PanzerInPrep}.

We consider the free commutative $\mathbb{Q}$-algebra generated by
a sole generator in degree zero, $\One$, which serves as a unit for the algebra. In degree three we put
\[ \Gamma_3=\wthree,\] whilst the only generator in degree four is
\[ \Gamma_4=\wfour. \]
In degree six we have $\Gamma_3\times\Gamma_3$ and
\[ \Gamma_{33}=\wtt,\] whilst in degree seven 
we have $\Gamma_3\times\Gamma_4$ and 
\[ \Gamma_{43}=\wft,\; \Gamma_{34a}=\wtfa,\; \Gamma_{34b}=\wtfb.\]
Note that $\Gamma_{34a},\Gamma_{34b}$ are the only two different topologies we can obtain by replacing  one of the five
vertices of $\Gamma_4$ by $\Gamma_3$. The four vertices of $\Gamma_4$ which are connected to an external momentum all give $\Gamma_{34b}$
(modulo permutations of edge labels), whilst inserting at the internal vertex of $\Gamma_4$ gives $\Gamma_{34a}$.
  
Finally, in degree eight we only consider  $\Gamma_4\times \Gamma_4$ and
insertion at the internal vertex:
\[ \Gamma_{44}=\wff.\]
At higher degrees, we only allow products of the generators listed so far.

We make this into a bi-algebra by setting $\Delta(\One)=\One\otimes \One$ and
$$ \Delta \Gamma_3=\Gamma_3\otimes \One + \One \otimes \Gamma_3,$$
$$ \Delta \Gamma_4=\Gamma_4\otimes \One + \One \otimes \Gamma_4,$$
$$ \Delta \Gamma_{33}=\Gamma_{33}\otimes \One + \One \otimes \Gamma_{33}+\Gamma_3\otimes \Gamma_3,$$
$$ \Delta \Gamma_{44}=\Gamma_{44}\otimes \One + \One \otimes \Gamma_{44}+\Gamma_4\otimes \Gamma_4,$$
$$ \Delta \Gamma_{43}=\Gamma_{43}\otimes \One + \One \otimes \Gamma_{43}+\Gamma_4\otimes \Gamma_3,$$
$$ \Delta \Gamma_{34a}=\Gamma_{34a}\otimes \One + \One \otimes \Gamma_{34a}+\Gamma_3\otimes \Gamma_4,$$
$$ \Delta \Gamma_{34b}=\Gamma_{34b}\otimes \One + \One \otimes \Gamma_{34b}+\Gamma_3\otimes \Gamma_4,$$
and $\Delta(h_1\times h_2)=\Delta(h_1)\times\Delta(h_2)$.

We get a Hopf algebra by setting $S(\One)=\One$, and recursively $S(h)=-m_H(S\otimes P)\Delta$, with $P$ the projection onto elements of positive degree,
i.e.\ the augmentation ideal.

Define two maps into the augmentation ideal 
 \[ B_+^3: H\to PH,\; \mathrm{and}\; B_+^4: H\to PH\] by
 \[ B_+^3(\One)=\Gamma_3,\; B_+^4(\One)=\Gamma_4,\]
 \[ B_+^3(\Gamma_3)=\Gamma_{33}, \;B_+^3(\Gamma_4)=\Gamma_{43},\]
 \[B_+^4(\Gamma_4)=\Gamma_{44},\;
 B_+^4(\Gamma_3)=\frac{1}{2}\left(\Gamma_{34a}+\Gamma_{34b}\right),\]
 and $B_+^i(h)=0,\,i\in \{3,4\}$, else.
 
 Then $\forall h\in \{\One,\Gamma_3,\Gamma_4\}$ and $i\in \{3,4\}$, 
 \[ \Delta B_+^i(h)=B_+^i(h)\otimes \One+(\mathrm{id}\otimes B_+^i)\Delta(h),
 \]
 which ensures that these maps behave as Hochschild one-cocycles in the examples below.
\begin{rem}
Effectively, we are working in a Hopf algebra of graphs generated and co-generated by $\Gamma_3$ and $\Gamma_4$, a quotient of the full Hopf algebra of graphs.
Note that, for example,
\[
\Delta B_+^4(\Gamma_{44})=0\not= B_+^4(\Gamma_{44})\otimes\One+(\mathrm{id}\otimes B_+^4)\Delta(\Gamma_{44})=\Gamma_4\otimes \Gamma_{44}.
\]
This is a consequence of restricting to a finite Hopf algebra. It poses no problems for our applications below in this finite Hopf algebra.
\end{rem} 
 
Now, let $\alpha_i:H\to\mathbb{C}$ be algebra maps, and let
$ b\alpha_i: H\to H$ be defined by
$$ b\alpha_i(h)=m(\mathrm{id}\otimes \alpha_i)\Delta(h)-\alpha_i(h)\One.$$
Then $b\alpha_i(\One)=0$ and $b\alpha_i(\Gamma_{j})=\alpha_i(\One)\Gamma_{j}$, $\forall i,j\in \{3,4\}$.
\begin{rem}
Were the $B_+^i$ to provide Hochschild one-cocycles, the $b\alpha_i$ would provide co-boundaries.
\end{rem}
\begin{rem}
We choose wheels in wheels as an example as results for them are on the brink of computability at the moment.
The methods of Francis Brown \cite{Brown1,Brown2} combined with \cite{BrKr1} allow to compute the period provided by $\Gamma_{33}$
\cite{PanzerBingen,PanzerInPrep}, whilst the periods from a symmetric combination of graphs $s_{34}$ as defined below are realistically in reach by this approch -and this approach only, it seems-, and the eight-loop $\Gamma_{44}$ period remains a challenge. 
\end{rem}
\subsection{Co-radical filtration}
Note that there is an obvious co-radical filtration and associated grading here, given by the kernels of $\sigma^{\star j}$, with $\sigma:=S\star P=m(S\otimes P)\Delta$, i.e.\ using projections into the augmentation ideal
combined with the co-product (see \cite{BrKr1}). 

We find in grading one the primitives $\Gamma_3,\Gamma_4$ and, more  interestingly, the primitive elements
\[ p_{33}:=2\Gamma_{33}-\Gamma_3\times\Gamma_3,\; p_{44}:=2\Gamma_{44}-\Gamma_4\times\Gamma_4,\]
and in particular 
\[p_{34}:=\overbrace{\frac{1}{2}\Gamma_{34a}+\frac{1}{2}\Gamma_{34b}+\Gamma_{43}}^{=:s_{34}}-\Gamma_3\times\Gamma_4,\]
which also defines the co-symmetric $s_{34}$ and 
\[ p_{a-b}:=\Gamma_{34a}-\Gamma_{34b}.\] 
They are all linear combinations of elements in filtration two which combine to give primitive
elements in the Hopf algebra, hence of co-radical degree 1.
Note that $s_{34}$ is a co-symmetric element (of co-radical degree two) in the Hopf algebra, which is the reason why
we can subtract the commutative product $\Gamma_3\times \Gamma_4$ to obtain a primitive element.

Let us also define a co-antisymmetric element in degree two,
\[ c_{34}=\frac{1}{2}\Gamma_{34a}+\frac{1}{2}\Gamma_{34b}-\Gamma_{43}.\]

Then, its reduced co-product $\Delta^\prime :=(P\otimes P)\Delta$ delivers
\[ \Delta^\prime c_{34}=\Gamma_3\otimes\Gamma_4-\Gamma_4\otimes\Gamma_3,\]
an element which indeed changes sign when we swap the elements on the lhs and rhs of the tensor product, contrary to
\[ \Delta^\prime s_{34}=\Gamma_3\otimes\Gamma_4+\Gamma_4\otimes\Gamma_3.\]

\subsection{Lie algebra aspects}
Let us now consider the Lie algebra $\mathcal{L}$ with generators $Z$ which are Kronecker-dual to the Hopf algebra generators $h$.
Its bracket is determined by
\[
\langle Z_i\otimes Z_j-Z_j\otimes Z_i,\Delta(h)\rangle =\langle [Z_i,Z_j],h\rangle,\,h\in H.
\]
Here, $\langle Z_a,\Gamma_b\rangle=\delta_{ab}$ is the Kronecker pairing between elements $Z_a\in \mathcal{L}$ and $\Gamma_b\in H$, and $a,b$ range over the set of subscripts $3,4,33,44,34a,\ldots$ used to denote
the graphs.

Consider also the corresponding universal enveloping algebra 
\[ U(\mathcal{L})=\mathbb{Q}\One\oplus \mathcal{L}\oplus (\mathcal{L}\otimes_S \mathcal{L}) \oplus \cdots .\]
Here, $\otimes_S$ denotes the symmetrized tensor-product, and $U(\mathcal{L})$ can be identified, albeit non-canonically, with the symmetric
tensor algebra of $\mathcal{L}$.

The Lie algebra $\mathcal{L}$ itself has a (descending) lower central series decomposition:
\[
\mathcal{L}_1:=\mathcal{L}, \mathcal{L}_k:=[\mathcal{L},\mathcal{L}_{k-1}],\, k>1.
\]

The co-product of an element in $H$ is not  co-commutative. It pays to decompose images of $\Delta^\prime$ and its iterations into symmetric and antisymmetric parts.
 
The idea on which we elaborate in the following is to map elements in the Hopf algebra to elements in the above universal enveloping algebra of its dual Lie algebra,
taking some extra information from physics: we will soon see that Feynman rules assign to Hopf algebra elements polynomials in a variable $L$, bounded by the co-radical degree,  which respects a decomposition 
into co-symmetric and co-antisymmetric terms in the Hopf algebra which is particularly illuminating in comparison with the universal enveloping algebra.

Concretely, let us consider the following map (extended by linearity)  $\sigma:\,H\to U(\mathcal{L})$.
We start with primitive elements $\Gamma_3,\Gamma_4,p_{33},p_{44},p_{34},p_{a-b}$, which, as we will see, all evaluate under the Feynman rules to terms linear in $L$:
\[
\sigma(\Gamma_i)=Z_i\in \mathcal{L}_1\subset U(\mathcal{L}),\,i\in 3,4,\]
\[
\sigma(p_{ii})=\overbrace{Z_{ii}}^{\in \mathcal{L}_1}\, i\in 3,4,
\]
\[
\sigma(s_{34})=\overbrace{Z_{s_{34}}}^{\in \mathcal{L}_1},
\]
\[
\sigma(p_{a-b})=\overbrace{Z_{p_{a-b}}}^{\in \mathcal{L}_1}.
\]
Note that under $\sigma$ these primitives have images $\in \mathcal{L}_1$, but $\not\in \mathcal{L}_2$.

Let us now consider non-primitive elements. As we will see under the Feynman rules, 
the next two examples give polynomials quadratic in $L$. This is reflected in $\sigma$:

\[
\sigma(\Gamma_{ii})=\overbrace{Z_{ii}}^{\in \mathcal{L}_1}
+\overbrace{\frac{1}{2}Z_i\otimes Z_i}^{\in \mathcal{L}_1\otimes_S \mathcal{L}_1},\, i\in 3,4,
\]
\[
\sigma(s_{34})=\overbrace{Z_{s_{34}}}^{\in \mathcal{L}_1}
+\overbrace{Z_3\otimes Z_4+Z_4\otimes Z_3}^{ \in \mathcal{L}_1\otimes_S \mathcal{L}_1},
\]
Note that the second symmetric tensor power shows up here, reflecting the $L^2$ term in the Feynman rules.

Finally, we have the co-antisymmetric element. It is of co-radical degree two, but is linear in $L$ under the Feynman rules. We map 
\[
\sigma(c_{34})=[Z_3,Z_4]\in \mathcal{L}_2,\;[Z_3,Z_4]\not\in\mathcal{L}_3,
\]
with
\[[Z_3,Z_4]=\frac{1}{2} Z_{34a}+\frac{1}{2}Z_{34b}-Z_{43}.\]
Note that the second symmetric tensor power does not show up here due to the co-antisymmetry of $c_{34}$. Nicely, the Feynman rules play along.

All others evaluations of $\sigma$ follow by linearity.
\begin{rem}
The fact that the Dynkin operator $S\star Y=m(S\otimes Y)\Delta$, -with $Y$ the grading operator multiplying a Hopf algebra element of co-radical degree $k$ by $k$-, of $H$ vanishes on products very much suggests to 
construct $\sigma$ as above. The fact that it maps pre-images $\sigma^{-1}$ of co-symmetric elements in $\mathcal{L}_1$ to primitive elements of $H$ motivates to look at the lower central series of
$\mathcal{L}$ for the co-antisymmetric elements. Also, note  that pre-images of co-symmetric elements can be generated from $\One$ through shuffles of one-cocycles, for example $(B_+^3B_+^4+B_+^4B_+^3)(\One)=s_{34}$.
\end{rem}

\section{Feynman Rules}
We now give the Feynman rules for Hopf algebra elements, next study them in examples provided by our small
Hopf algebra, and discuss the induced Feynman rules on the Lie side at the end.

Feynman rules on the Hopf algebra side are provided for scalar fields from the two Symanzik polynomials, together with the above Hopf algebra structure. For gauge fields, a third polynomial \cite{Cor} allows to obtain the Feynman rules for gauge theory from the scalar field rules \cite{KrSarsWvS}.
We follow \cite{BrKr1,BrKr2,KrSarsWvS}.
 \subsection{The first Kirchhoff polynomial $\psi_\Gamma$}
For the first Kirchhoff polynomial consider the short exact sequence
\begin{equation}
0\to H^1\to\mathbb{Q}^E\overbrace{\to}^{\partial}\mathbb{Q}^{V,0}\to0.
\end{equation}
Here, $H^1$ is provided by a chosen basis for the independent loops of a graph $\Gamma$.
 $E=|E^\Gamma|$ is the number of edges and 
$V=|V^\Gamma|$ the number of vertices, so $\mathbb{Q}^E$ is an $E$-dimensional $\mathbb{Q}$-vectorspace generated by the edges, similar $\mathbb{Q}^{V,0}$
for the vertices with a side constraint setting the sum of all vertices to zero.

Consider the matrix (see \cite{BEK,BlKr})
\begin{equation*}
N_0\equiv (N_0)_{ij}=\sum_{e\in l_i\cap l_j}A_e,
\end{equation*}
for $l_i,l_j\in H^1$.

Define the {\it first Kirchhoff polynomial} as the determinant
\begin{equation*}
\psi_\Gamma:=|N_0|.
\end{equation*}
\begin{prop}(\cite{Kirchhoff}, see also \cite[Prop.2.2]{BEK})
The first Kirchhoff polynomial can be written as
$$\psi_\Gamma=\sum_{T}\prod_{e\not\in T}A_e$$
where the sum on the right is over spanning trees $T$ of $\Gamma$.
\end{prop}

\subsection{The second Kirchhoff polynomial $\phi_\Gamma$ and $|N|_{\mathrm{Pf}}$}
To each edge $e$ we assign an auxiliary  four-vector $\xi_e$.

Let then $\sigma^i$, $i\in 1,2,3$ be the three Pauli matrices, and $\sigma^0=\One_{2\times 2}$ the unit matrix.

For the second Kirchhoff polynomial,
augment the matrix $N_0$ to a new matrix $N$ in the following way:
\begin{enumerate}
\item Assign to each edge $e$ a quaternion \[
\mathbf{q}_e:=\xi_{e,0}\sigma^0-i\sum_{j=1}^3 \xi_{e,j}\sigma^j,\]
so that  $\xi_e^2\One_{2\times 2}=\mathbf{q}_e\bar{\mathbf{q}}_e$,
 and to the loop $l_i$,
the quaternion \[
u_i=\sum_{e\in l_i} A_e \mathbf{q}_e.\]\\
\item Consider the column vector $u=(u_i)$ and the conjugated transposed row vector 
$\bar{u}$. Augment $u$ as the rightmost column vector to $M$, and $\bar{u}$ as the bottom row vector.
\item Add a new diagonal entry at the bottom right  $ \sum_e \mathbf{q}_e\bar{\mathbf{q}}_e A_e$.
\end{enumerate}

Note that by momentum conservation, to each vertex, we assign a momentum $\xi_v$,
and a corresponding quaternion $\mathbf{q}_v$.
\begin{rem}
Note that we use that we work in four dimensions of space-time, by rewriting the momentum four-vectors in a quaternionic basis.
This strictly four-dimensional approach can be extended to twistors \cite{BlochTwistors}.
\end{rem}
The matrix $N$ has a well-defined Pfaffian determinant (see \cite{BlKr})
with a remarkable form obtained for generic $\xi_e$ and hence generic $\xi_v$:
\begin{lem}(\cite[Eq.3.12]{BlKr})\label{gensym}
$$|N|_{\mathrm{Pf}}=-\sum_{T_1\cup T_2}\left(\sum_{e\not\in T_1\cup T_2} \tau(e)\xi_e\right)^2\prod_{e\not\in T_1\cup T_2}A_e,$$
where $\tau(e)$ is $+1$ if $e$ is oriented from $T_1$ to $T_2$ and $-1$ else.
\end{lem} 
Here, $T_1,T_2$ are two trees such that their union contains all vertices of the graph, i.e.\ $T_1\cup T_2$ is a spanning 2-tree.
 
Note that $|N|_{\mathrm{Pf}}=|N|_{\mathrm{Pf}}(\{\xi_v\})$ is a function of all $\xi_v$, $v\in \Gamma^{[0]}$.
From the view-point of graph theory, this is the natural polynomial. It gives the second Symanzik polynomial upon setting the $\xi_e$ in accordance with the external momenta
$p_e$:
\[
Q: \xi_e\to +p_e.
\] 
\begin{rem}
Adding to the second Symanzik polynomial a term $\psi_\Gamma\sum_{e\in \Gamma} A_em_e^2$ allows to treat masses $m_e$.
\end{rem}
\begin{rem}
For $\gamma\subset\Gamma$ a non-trivial subgraph, and $\kappa\in\{\phi,\psi\}$ we have almost factorization: $\kappa_\Gamma=\kappa_{\Gamma/\gamma}\psi_\gamma+R^\kappa_{\Gamma,\gamma}$, with the remainders
$R^\kappa_{\Gamma,\gamma}$ homogeneous polynomials of higher degree in the sub-graph variables than $\psi_\gamma$.  
\end{rem}
\subsection{The unrenormalized integrand}
In Schwinger parametric form, the unrenormalized  Feynman amplitude $\mathcal{I}_\Gamma$ (omitting trivial overall factors of powers of $\pi$ and such)
comes from an integrand $I_\Gamma$
\begin{equation}\label{param}
\mathcal{I}_\Gamma = \int \underbrace{\frac{e^{-\frac{\phi_\Gamma}{\psi_\Gamma}}}{\psi_\Gamma^2}}_{I_\Gamma}\prod_e d\! A_e.
\end{equation}
This form gets modified if we allow for spin and other such complications. An exhaustive study of how to obtain gauge theory amplitudes from such an integrand is given in \cite{KrSarsWvS}.
\begin{rem}
A regularized integrand can be obtained by raising the denominator $1/\psi^2_\Gamma$ to a noninteger power (dimensional regularization), or 
multiplication by non-integer powers of edge variables, together with suitable $\Gamma$-functions (analytic regularization). The latter suffices to treat the Mellin transforms 
as used for example in \cite{YeatsMadrid}
and discussed below.
\end{rem}
\subsection{The renormalized integrand}
We can render the integrand  $I_\Gamma$ integrable wrt to the domain $\sigma_\Gamma$ prescribed by parametric integration by a suitable sum over forests. We define
\[
I^R_\Gamma:=\sum_{f\in\mathcal{F}_\Gamma}(-1)^{|f|} I_{\Gamma/f} I_f^0,
\]
where for $f=\bigcup_i\gamma_i$, $I_f=\prod_i I_{\gamma_i}$ and the superscript ${}^0$ indicates that kinematic variables are specified according to renormalization conditions. 

The formula for $I_\Gamma^R$ is correct as long as all sub-graphs are overall log-divergent, the necessary correction terms in the general case are given in \cite{BrKr1}. In our examples below, we can always identify the one log-divergent subgraph -if any- with the unique non-trivial forest.

\subsection{The renormalized integral}
Accompanying this integrand is the renormalized result which can be written projectively:
\[
\Phi^R(\Gamma):=\int_{\mathbb{P}^{|E^\Gamma_I|}(\mathbb{R}_+)}\sum_{f\in\mathcal{F}_\Gamma}(-1)^{|f|} \frac{\ln{\frac{\phi_{\Gamma/f}\psi_f+\phi^0_f\psi_{\Gamma/f}}{\phi_{\Gamma/f}^0\psi_f+\phi^0_f\psi_{\Gamma/f}}}}{\psi^2_{\Gamma/f}\psi^2_f}\Omega_\Gamma,
\]
for notation see \cite{BrKr1,BrKr2} or \cite{KrSarsWvS}. Let us just mention that for the domain of integration we will abbreviate from now on
\[
\mathbb{P}^{|E^\Gamma_I|}(\mathbb{R}_+)=\mathbb{P}_\Gamma.
\] 
Note that this is a well-defined integral obtained from the use of the forest formula.
It is obtained without using an intermediate regulator. It is well-suited to analyse the mathematical structure
of perturbative contributions to Green functions. 

Also, combining this approach
with \cite{KrSarsWvS}, it furnishes a reference point against which to check in a situation where intermediate regulators are spoiling the symmetries of the theory. 

Below, we will shortly compare the structure of this integrand to the appaearance of analytic regulators provided
by anomalous dimensions of quantum fields, wich then define Mellin transforms for the primitives in the Hopf algebra.

\subsection{Scales and Angles}
Feynman graphs have their external edges labelled by momenta, and internal edges labelled by masses. 

Renormalized Feynman rules above are therefore functions of scalar products $Q_i\cdot Q_j$ and mass-squares $m_e^2$.
Equivalently, upon defining a positive definite  linear combination $S$ of such variables, we can write them as functions of such a scale $S$, and angles
$\Theta_{ij}:=Q_i\cdot Q_j/S$, $\Theta_e:=m_e^2/S$. We use $S^0,\Theta^0_{ij},\Theta^0_e$ to specify scale and angles for
the renormalization point. A graph which furnishes only a single scalar
product $Q\cdot Q$ as a scale is denoted a 1-scale graph.

Isolating short-distance singularities in 1-scale sub-graphs has many advantages, including a systematic separation of angles and scales, and a clean approach to the renormalization group as well as an identification of the freedom provided by exact terms in the Hochschild cohomology,
as we discuss below, see also \cite{PanMas,PanMadrid,PanKr}.

Following \cite{BrKr1}, we have the decomposition
\begin{thm}
\[
\Phi^R(S/S^0,\{\Theta,\Theta^0\})=\Phi_{\mathrm{fin}}^{-1}(\{\Theta^0\})\star\Phi_{\mathrm{\makebox{1-s}}}^R(S/S^0)\star\Phi_{\mathrm{fin}}(\{\Theta\}).
\]
\end{thm}
Here, the angle-dependent Feynman rules $\Phi_{\mathrm{fin}}$ are computed by eliminating short-distance singularities through the comparison, via the Hopf algebra,
with 1-scale graphs evaluated at the same scale as the initial graphs, while the 1-scale Feynman rules $\Phi_{\mathrm{\makebox{1-s}}}^R(S/S_0)$ eliminate short-distance singularities by renormalizing 1-scale graphs at a reference scale $S_0$. 
\begin{rem}
Feynman rules in parametric renormalization allow to treat the computation of Feynman graphs as a problem of algebraic geometry, analysing 
the structure of two kinds of homogeneous polynomials \cite{Brown1,Brown2,BrKr1,BogLueMadrid}.
\end{rem}
\begin{rem}
The fact that it is basically the denominator structure which determines the computability of Feynman graphs in parametric renormalization
makes this approach very efficient in computing periods in the $\epsilon$-expansion of regularized integrands.
\end{rem}
\begin{rem}
We assume throughout that angles and scales are such that we are off any infrared singularities, for example by off-shell external momenta.
The latter would not be cured by the forest sums which eliminate short-distance singularities. 
\end{rem}
\section{Examples} 
\subsection{Overall finite graphs}
From now on, we write $\phi_\Gamma=\phi_\Gamma(\Theta),\phi_\Gamma^0\equiv\phi_\Gamma(\Theta_0)$. For a 1-scale graph $\Gamma$,
we let $\Gamma^\bullet$ be the graph where the two external vertices of $\Gamma$ are identified. One has $\phi_\Gamma=\psi_{\Gamma^\bullet}$. 

Assume we are considering a superficially convergent graph $\Gamma$
free of subdivergences.
For example, a graph $\Gamma$  in four dimensions of space time on $n>2|\Gamma|$ edges delivers the integrable form
\[
\frac{1}{S^{n-2|\Gamma|}}\int \frac{1}{\psi^2}\left(\frac{\psi}{\phi(\Theta)}\right)^{n-2|\Gamma|}\Omega_\Gamma.
\]
This is polynomial in the scale dependence, while the angle dependence is di-logarithmic for good reasons \cite{BlKr}.

Inserting logarithmic subdivergences,
we get the integrable form (it is integrable as long as external momenta are off-shell such that no infrared singularities arise)
\[
\frac{1}{S^{n-2|\Gamma|}}\int_{\mathbb{P}_\Gamma} \sum_f (-1)^{|f|}\frac{1}{\psi^2_{\Gamma/f}\psi^2_{f}}\left(\frac{\psi_{\Gamma/f}\psi_{f}}{\phi_{\Gamma/f}\psi_f+\phi_f^0\psi_{\Gamma/f}}\right)^{n-2|\Gamma|}\Omega_\Gamma.
\]
Note that $\phi_\emptyset=0,\psi_\emptyset=1$.
\begin{rem}
Note that for the logarithmic divergent case $n=2|\Gamma|$, we got a logarithm in the numerator of the renormalized integrand, 
reflecting the superficial degree of divergence zero. 
In the convergent case, the above power of $n-2|\Gamma|$ is then reflecting the superficial degree of convergence $2(n-2|\Gamma|)$. 
\end{rem}
\subsection{Primitive graphs}
Consider now a logarithmically divergent graph without sub-divergences, $L=\ln S/S^0$.
Then,
$$\Phi_R(\Gamma)= c_\Gamma^1 L+c_\Gamma^0(\Theta,\Theta_0).$$
We have
$$c_\Gamma^1=\int_{\mathbb{P}(\Gamma)} \frac{\Omega_\Gamma}{\psi_\Gamma^2}, $$
$$c_\Gamma^0(\Theta,\Theta_0)=\int_{\mathbb{P}(\Gamma)} \frac{\ln{\frac{\phi_\Gamma}{\phi_\Gamma^0}}\Omega_\Gamma}{\psi_\Gamma^2}.$$
The finite part $c_\Gamma^0(\Theta,\Theta_0)$ can equivalently be expressed in the form of overall finite graphs.
Let $P_e$ be the propagator at edge $e$, $P_e^0$ the same propagator, but with its external momenta evaluated as prescribed by the renormalization condition.
Then,
\[
\frac{1}{P_e}-\frac{1}{P_e^0}=\frac{P_e^0-P_e}{P_eP_e^0},
\]
where internal loop momenta in edge $e$ drop out in the difference $P_e^0-P_e$.
 
By telescoping we can extend to products of propagators provided by graphs, and hence express the finite part of an overall logarithmically divergent graph as an overall convergent graph, which is an element of a larger Hopf algebra provided by general Feynman integrals.

\subsection{Structure of a graph with a sub-divergence}
Consider $\Gamma=\Gamma_{43}$ say, as a generic example. We have $\Delta^\prime(\Gamma)=\Gamma_4\otimes\Gamma_3$.

Then $$\Phi_R(\Gamma)= c_\Gamma^2 L^2+c_\Gamma^1(\Theta,\Theta_0)L+c_\Gamma^0(\Theta,\Theta_0).$$
We have
$$\Phi_R(\Gamma_{43})= \int_{\mathbb{P}_\Gamma}\left( \frac{\ln{\frac{\frac{S}{S_0}\phi_\Gamma}{\phi_\Gamma^0}}}{\psi_\Gamma^2}
-\frac{
\ln{\frac{\frac{S}{S_0}\phi_{\Gamma_3}\psi_{\Gamma_4}+\phi_{\Gamma_4}^0\psi_{\Gamma_3}}{\phi_{\Gamma_3}^0\psi_{\Gamma_4}+\phi_{\Gamma_4}^0\psi_{\Gamma_3}}
}}{\psi_{\Gamma_4}^2\psi_{\Gamma_3}^2}\right)\Omega_\Gamma.$$

We then have for the scale independent part
$$c_\Gamma^0(\Theta,\Theta_0)= 
 \int_{\mathbb{P}_\Gamma}\left( \frac{\ln{\frac{\phi_\Gamma}{\phi_\Gamma^0}}}{\psi_\Gamma^2}
-\frac{
\ln{\frac{\phi_{\Gamma_3}\psi_{\Gamma_4}+\phi_{\Gamma_4}^0\psi_{\Gamma_3}}{\phi_{\Gamma_3}^0\psi_{\Gamma_4}+\phi_{\Gamma_4}^0\psi_{\Gamma_3}}
}}{\psi_{\Gamma_4}^2\psi_{\Gamma_3}^2}\right)\Omega_\Gamma,
$$
and for the term linear in $L$:
\begin{equation}\label{ainb}
c_\Gamma^1(\Theta,\Theta_0)= 
 \int_{\mathbb{P}_\Gamma}\left( \frac{1}{\psi_\Gamma^2}
-\frac{\phi_{\Gamma_3}\psi_{\Gamma_4}}{\psi_{\Gamma_4}^2\psi_{\Gamma_3}^2\left[\phi_{\Gamma_3}\psi_{\Gamma_4}+\phi_{\Gamma_4}^0\psi_{\Gamma_3} \right]}\right)\Omega_\Gamma.
\end{equation}
The term quadratic in $L$ gives
$$c_\Gamma^2= 
 \int_{\mathbb{P}_\Gamma}\left( \frac{\phi_{\Gamma_3}\psi_{\Gamma_4}\psi_{\Gamma_3}\phi_{\Gamma_4}^0}{\psi_{\Gamma_4}^2\psi_{\Gamma_3}^2\left[\phi_{\Gamma_3}\psi_{\Gamma_4}+\phi_{\Gamma_4}^0\psi_{\Gamma_3} \right]^2}\right)\Omega_\Gamma.
$$
Scaling out from the edge variables of the subgraph one of its variables $\lambda$ say- and integrating it, so that $\Omega_\Gamma\to \Omega_{\Gamma_3}\wedge\Omega_{\Gamma_4}\wedge d\lambda$
(a careful treatment of such changes of variables is in \cite{BrKr1})
gives us 
\begin{eqnarray*}
c_\Gamma^2 & = & 
 \int_{\mathbb{P}_{\Gamma_3}\times \mathbb{P}_{\Gamma_4}}\left(\int_0^\infty  \frac{\phi_{\Gamma_3}(\Theta)\psi_{\Gamma_4}\psi_{\Gamma_3}\phi_{\Gamma_4}(\Theta_0)}{\psi_{\Gamma_4}^2\psi_{\Gamma_3}^2\left[\phi_{\Gamma_3}(\Theta)\psi_{\Gamma_4}+\lambda\phi_{\Gamma_4}(\Theta_0)\psi_{\Gamma_3} \right]^2}d\lambda\right) \Omega_{\Gamma_3}
 \wedge\Omega_{\Gamma_4}\\
 & = &
\int_{\mathbb{P}_{\Gamma_3}}\frac{1}{\psi_{\Gamma_3}^2}\Omega_{\Gamma_3}\int_{\mathbb{P}_{\Gamma_4}} \frac{1}{\psi_{\Gamma_4}^2}
\Omega_{\Gamma_4},
\end{eqnarray*}
which fully exhibits the desired factorization.

One easily checks that $\partial_L^k$ vanishes for $k$ greater than the co-radical degree.

\subsection{Periods from insertion places}
Let us now consider the primitive $p_{a-b}$. The two graphs involved are distinguished only by the insertion place into which we insert the subgraph $\Gamma_3$. From the previous result is it evident that for $p_{a-b}$ we could at most  find up to a linear term in $L$\[
\Phi_R(p_{a-b})=c_{p_{a-b}}^1 L+c_{p_{a-b}}^0(\Theta,\Theta^0).
\]
We find for this   scheme-independent -and hence well-defined- period
$$c_{p_{a-b}}^1=\int_{\mathbb{P}_{\Gamma_{34}}} \left( \frac{1}{\psi_{\Gamma_{34a}}^2}-\frac{1}{\psi_{\Gamma_{34b}}^2}\right) \Omega_{\Gamma_{34}}, $$ 
where $\mathbb{P}_{\Gamma_{34}}$ and $\Omega_{\Gamma_{34}}$ are obviously independent of the insertion place.

Note that the difference is completely governed by $R^\psi_{\Gamma_{34a},\Gamma_3}$ as compared to 
$R^\psi_{\Gamma_{34b},\Gamma_3}$, while for the term constant in $L$ we also need to consider
$R^\phi_{\Gamma_{34a},\Gamma_3}$ as compared to 
$R^\phi_{\Gamma_{34b},\Gamma_3}$.

From now on we discard the constant 
terms in $L$, which we regard as originating from overall convergent integrals.

\subsection{Periods for co-commutative elements}
Next, let us look at $s_{34}$ which is of co-radical degree 1. Clearly, 
\[
\Phi_R(s_{34})=c_3^1c_4^1 L^2+c_{s_{34}}^1.
\]
In general, $c_{s_{34}}^1$ is not a period but rather a complicated function of $\Theta,\Theta^0$.

We now assume that we subtract at $\Theta=\Theta_0$. 

$c_{s_{34}}^1$ could then still be a function of the angles $\Theta$.
Instead, it is a constant, as is immediately clear by using Eq.(\ref{ainb}). This constant is a period which hopefully is known to us soon enough
using the methods of \cite{PanzerInPrep}. 
\begin{eqnarray}
c_{s_{34}}^1 & = &  
 \int_{\mathbb{P}_\Gamma}\left( 
 \frac{1}{2} \frac{1}{\psi_{\Gamma_{34a}}^2}+\frac{1}{2} \frac{1}{\psi_{\Gamma_{34b}}^2}+\frac{1}{\psi_{\Gamma_{43}}^2}\right.\nonumber\\
  & & \left. 
-\frac{\phi_{\Gamma_3}\psi_{\Gamma_4}+\phi_{\Gamma_4}\psi_{\Gamma_3}}{\psi_{\Gamma_4}^2\psi_{\Gamma_3}^2\left[\phi_{\Gamma_3}\psi_{\Gamma_4}+\phi_{\Gamma_4}\psi_{\Gamma_3} \right]}\right)\Omega_\Gamma\nonumber\\ & = &
 \int_{\mathbb{P}_\Gamma}\left( 
 \frac{1}{2} \frac{1}{\psi_{\Gamma_{34a}}^2}+\frac{1}{2} \frac{1}{\psi_{\Gamma_{34b}}^2}+ \frac{1}{\psi_{\Gamma_{43}}^2}-\frac{1}{\psi_{\Gamma_4}^2\psi_{\Gamma_3}^2}\right) \Omega_\Gamma,
\end{eqnarray}
where the notation $\mathbb{P}_\Gamma,\Omega_\Gamma$ is justified, as edges can be consistently labeled in all terms.
Note that the step above from the first to the second line follows as we have $\phi_{\Gamma}(\Theta_0)=\phi_{\Gamma}(\Theta)=\phi_\Gamma,\forall \Gamma$.

\subsection{Angle dependence in commutators}
For anti-cocommutative elements like $c_{34}$ angle dependence remains, even if we set $\Theta=\Theta^0$. In such a setting, we find 
\[\Phi_R(c_{34})=c_{34}^1(\Theta)L.\]
\begin{eqnarray}
c_{34}^1(\Theta) & = &  
 \int_{\mathbb{P}_\Gamma}\left( 
 \frac{1}{2} \frac{1}{\psi_{\Gamma_{34a}}^2}+\frac{1}{2}\frac{1}{\psi_{\Gamma_{34b}}^2}-\frac{1}{\psi_{\Gamma_{43}}^2}\right.\nonumber\\
  & & \left. 
-\frac{\phi_{\Gamma_4}\psi_{\Gamma_3}-\phi_{\Gamma_3}\psi_{\Gamma_4}}{\psi_{\Gamma_4}^2\psi_{\Gamma_3}^2\left[\phi_{\Gamma_3}\psi_{\Gamma_4}+\phi_{\Gamma_4}\psi_{\Gamma_3} \right]}\right)\Omega_\Gamma.
\end{eqnarray}
In this way, when renormalizing at unchanged scattering angles, angle dependence is relegated to anti-cocommutativity.
\subsection{1-scale sub-graphs vs $\Phi_{1-s}$}
In \cite{BrKr1} scale and angle depenence were separated using  1-scale renormalized Feynman rules $\Phi_{1-s}^R$.
These are massless Feynman rules which act by choosing two distinct vertices for each subdivergent graph $\gamma\subset\Gamma$ and evaluating the counterterms for this subgraph
treating it as a 1-scale graph $\gamma_2$, so that we have $\phi_\gamma=\psi_{\gamma^\bullet}$. 

Also, $\Gamma$ itself allows external momenta only at two distinct vertices.

As discussed in \cite{BrKr1}, see also \cite{BrKr2} for a detailed example, we can enlarge the set of graphs to be considered by graphs $G_2$ say so that
$\Gamma/\gamma=G_2/g_2$, and \[\phi_{\Gamma/\gamma}\psi_{g_2}+\psi_{\gamma^\bullet}\psi_{\Gamma/\gamma}\] is the two-vertex join
of co- and subgraph \cite{BrKr1}. In $G_2$, edges which connect $G_2-g_2$ to $g_2$ all originate from the two distinct vertices chosen in $\gamma$.
the reader will have to consider \cite{BrKr1} for precise definitions.

Feynman rules for graphs $G_2$ are the canonical $\Phi_R$ subtracting at $S=S_0$, and for example for $\Gamma=\Gamma_{43}$,
\[
\Phi_R(G_2)-\Phi_{1-s}^R(\Gamma)=\int_{\mathbb{P}_\Gamma}\left( \frac{1}{\psi^2_{G_2}}-\frac{1}{\psi^2_\Gamma}\right)\Omega_\Gamma,
\]
gives us another period.  

$G_2$ in this example is the graph
$$G_2= \gt,$$ with the understanding that momenta are zero at two of its four marked external vertices
when acted upon by $\Phi^R_{1-s}$. Its wheel with four spokes subgraph is rendered 1-scale upon enforcing a five-valent vertex and hence must be treated in a suitably enlarged Hopf algebra.

The general case is studied in \cite{BrKr1} in great detail.

\subsection{Mellin transforms and 1-scale subgraphs}
Graphs of the form $G_2$ can be computed by defining a suitable Mellin transform \cite{KrY,KrY6p,Ytheses}.
This holds even if the co-graph is not 1-scale, the important fact being that the subgraph is.

This Mellin transform is defined by raising a quadric $Q(e)$ for an internal edge $e$, or a linear combination of such quadrics,
to a non-integer power 
\[
\frac{1}{Q(e)}\to \frac{1}{Q(e)^{1+\rho}},
\] 
in a cograph which has no subdivergences.
This defines a Mellin transform (staying in the above example)  \[M_{\Gamma_3}(\rho,L)=e^{-\rho L}f_3(\rho),\] 
where $f_3(\rho)$ has a first order pole in $\rho$ at zero with residue $6\zeta(3)=c_{\Gamma_3}^1$.

Also, $f_3(\rho)=f_3(1-\rho)$.
We set 
\[
f_3(\rho)=\frac{6\zeta(3)}{\rho(1-\rho)}(1+d_3^1(\Theta)\rho+\mathcal{O}(\rho^2)).
\]

We can compute $\Phi_R(G_2)$ as
\[
\Phi_R(G_2)=-\overbrace{c_{\Gamma_4}^1}^{20\zeta(5)}\partial_\rho (e^{-\rho L}-1)f(\rho).
\]
One hence finds that $c_{G_2}^2=60\zeta(3)\zeta(5)$ and $c_{G_2}^1=120\zeta(3)\zeta(5)-\zeta(5)d_3^1(\Theta)$.
 
Can we confirm this structure from the parametric approach?

We first note that \[|\psi_{G_2/\Gamma_4}|=|\phi_{G_2/\Gamma_4}|-1,\] \[|\psi_{G_2/\Gamma_4}|=|\psi_{G_2-\Gamma_4}|+1.\] 

Returning to affine coordinates, scaling out a subgraph variable and integrating it, using $R^\psi_{G_2,\Gamma_4}=\psi_{\Gamma_4^\bullet}\psi_{G_2-\Gamma_4}$ and returning to $\mathbb{P}_{\Gamma_3}\times\mathbb{P}_{\Gamma_4}$
delivers
\begin{eqnarray}
c_{G_2}^1 =\underbrace{\int_{\mathbb{P}_{\Gamma_4}}\frac{1}{\psi_{\Gamma_4}^2} \Omega_{\Gamma_4}}_{20\zeta_5}
\underbrace{\int_{\mathbb{P}_{\Gamma_3}}\frac{1}{\psi_{\gamma_3}^2}\left( 1-\ln{\frac{\psi_{G_2-\Gamma_4}\phi_{\Gamma_3}}{\psi^2_{\Gamma_3}}}\right) \Omega_{\Gamma_3}}_{6\zeta_3-d_3^1},
\end{eqnarray}
as desired.
\subsection{Well-defined periods from the dual of commutators}

Finally, a further angle independent period is furnished by the 1-scale version $c_{34,1-s}$ of the anti-cocommutative $c_{34}$.  
\[
c_{c_{34,1-s}}^1=\int_{\mathbb{P}_\Gamma} \left( \frac{1}{\psi^2_{\Gamma_{43}}}
-\frac{1}{2\psi^2_{\Gamma_{34a}}}-\frac{1}{2\psi^2_{\Gamma_{34b}}}
-\frac{\psi_{\Gamma_3^\bullet}\psi_{\Gamma_4}-\psi_{\Gamma_4^\bullet}\psi_{\Gamma_3}}{\psi^2_{\Gamma_3}\psi^2_{\Gamma_4}(\psi_{\Gamma_3^\bullet}\psi_{\Gamma_4}+\psi_{\Gamma_4^\bullet}\psi_{\Gamma_3})}\right)  \Omega_\Gamma,
\]
where it is understood that all graphs have their subgraphs as 1-scale subgraphs as in $G_2$.
By the previous result, this can be decomposed into two separate projective integrals.

\subsection{The role of exact co-boundaries}
We have seen that when a subgraph $\gamma$ is 1-scale, the evaluation of the full graph factorizes the period $c_\gamma^1$.
This is the crucial fact which allows to use co-boundaries to alter the Taylor coefficients of Mellin transforms
\cite{PanKr,PanMas,PanMadrid}.

For example, with $B_+^3$ now effecting a 1-scale insertion, $B_+^3(\Gamma_4)=G_2$ and $b\alpha_3(\Gamma_4)=\alpha_3(\One)\Gamma_4$:
\[
\Phi_R((B_+^3+b\alpha_3)(\Gamma_4)=\Phi_R(\Gamma_2)+20\zeta(5)\alpha_3(\One)L,
\]
where we are free to choose $\alpha_3(\One)$ to modify $d_3^1(\Theta)$, a useful fact in light of the manipulations in \cite{Ytheses,YeatsMadrid}. 
\section{Feynman rules from a Lie viewpoint}
The map $\sigma:H\to U(\mathcal{L})$ can be combined with projectors $T_k$ into the $k$-th symmetric tensorpower of $\mathcal{L}$.
Let then \[\sigma_k:=T_k\circ \sigma.\] Then, $\sigma_1$ takes values in $\mathcal{L}$, $\sigma_2$ takes values in $\mathcal{L}\otimes_s\mathcal{L}$,
and so on.

The map $\sigma_1:H\to \mathcal{L}$ is such that an element $h\in H$ in the $k$-th co-radical filtration (so that $\Delta^{\prime k}(h)\not=0$)
has contributions  in $\mathcal{L}_k$ at most, for example the co-radical degree two $c_{34}$ fulfils this bound as it  maps to $[Z_3,Z_4]\in\mathcal{L}_2$. 

In general, a co-radical degree $k$ element  has a non-vanishing component in $\mathcal{L}_k$ if and only if 
$\Delta^{\prime k}(h)$ contains corresponding anti-symmetric elements.

The symmetric parts in $\Delta^{\prime k}(h)$ map under $\sigma_1$ to an element $l_1(h)\in \mathcal{L}_1$ say, $l_1(h)\not\in \mathcal{L}_2$, so that the Dynkin operator
$S\star Y$ maps the pre-image $\sigma_1^{-1}(l_1)$ to a primitive element in $H$, 
\[\Phi_R(S\star Y \sigma_1^{-1}(l_1(h)))=c_{l_1}^1 L,\]
while all other terms in $c_h^1$ come from the pre-images of elements in $\mathcal{L}_k,k>1$.

Finally, pre-images of $\sigma_k$  provide the contributions to order $L^k$ similarly, in full accordance with the co-radical filtration 
and the renormalization group \cite{BrKr1}.
 
\bibliographystyle{amsalpha}

\begin{thebibliography}{99}
\bibitem{BlochTwistors} S.\ Bloch, talk at Spring School: Feynman Graphs and Motives, Bingen, march 2013,
{\tt http://www.sfb45.de/events/spring-school-feynman-graphs-and-motives}.

\bibitem{BEK} S.\ Bloch, H.\ Esnault, D.\ Kreimer,
 {\em On Motives associated to graph polynomials,}
  Commun.\ Math.\ Phys.\  {\bf 267} (2006) 181
  [math/0510011 [math-ag]].
  %%CITATION = MATH/0510011;%%

\bibitem{BlKr} S.\ Bloch, D.\ Kreimer,
  {\em Feynman amplitudes and Landau singularities for 1-loop graphs,}
  Commun.\ Num.\ Theor.\ Phys.\  {\bf 4} (2010) 709
  [arXiv:1007.0338 [hep-th]].
  %%CITATION = ARXIV:1007.0338;%%


\bibitem{BogLueMadrid}
%\cite{Bogner:2013tia}
%\bibitem{Bogner:2013tia}
  C.~Bogner and M.~L\"uders,
  {\em Multiple polylogarithms and linearly reducible Feynman graphs,}
  arXiv:1302.6215 [hep-ph], these proceedings.
  %%CITATION = ARXIV:1302.6215;%%


\bibitem{Pisa}
%\cite{Broadhurst:1995km}
%\bibitem{Broadhurst:1995km}
  D.~J.~Broadhurst and D.~Kreimer,
  {\em Knots and numbers in $\Phi^4$ theory to 7 loops and beyond,}
  Int.\ J.\ Mod.\ Phys.\ C {\bf 6} (1995) 519
  [hep-ph/9504352].
  %%CITATION = HEP-PH/9504352;%%
  %68 citations counted in INSPIRE as of 12 Mar 2013

\bibitem{Brown1} F.\ Brown,
{\em On the periods of some Feynman integrals}, arXiv:0910.0114v1, (2009), 1-69.

\bibitem{Brown2} F. Brown, {\em The Massless higher-loop two point function}, Comm. in Math. Physics  287, Number 3, (2009), 925-958.

\bibitem{BrKr1}
%\cite{Brown:2011pj}
%\bibitem{Brown:2011pj} 
  F.~Brown and D.~Kreimer,
  {\em Angles, Scales and Parametric Renormalization,}
  Lett.\ Math.\ Phys.\  {\bf 103} (2013) 933-1007,
  [arXiv:1112.1180 [hep-th]].
  %ARXIV:1112.1180;%%
  %7 citations counted in INSPIRE as of 12 Mar 2013

\bibitem{BrKr2}
%\cite{Brown:2012cb}
%\bibitem{Brown:2012cb}
  F.~Brown and D.~Kreimer,
  {\em Decomposing Feynman rules,}
  arXiv:1212.3923 [hep-th], Proceedings of Science (LL2012)049.
  %%CITATION = ARXIV:1212.3923;%%

\bibitem{Kirchhoff}
G.~Kirchhoff, {\em \"Uber die Aufl\"osung der Gleichungen, auf welche man bei der Untersuchung
der linearen Vertheilung galvanischer Str\"ome gef\"uhrt wird}, Annalen der Physik und Chemie
72 no. 12 (1847), 497-508.




\bibitem{PanKr}
%\cite{Kreimer:2012nk}
%\bibitem{Kreimer:2012nk} 
  D.~Kreimer and E.~Panzer,
  {\em Renormalization and Mellin transforms,}
  arXiv:1207.6321 [hep-th], Proceedings Linz July 2012, to appear.
  %%CITATION = ARXIV:1207.6321;%%
  %1 citations counted in INSPIRE as of 12 Mar 2013

\bibitem{KrSarsWvS}
%\cite{Kreimer:2012jw}
%\bibitem{Kreimer:2012jw}
  D.~Kreimer, M.~Sars and W.~D.~van Suijlekom,
  {\em Quantization of gauge fields, graph polynomials and graph cohomology,}
  Annals Phys.\  {\bf 336} (2013) 180-222
  [arXiv:1208.6477 [hep-th]].
  %%CITATION = ARXIV:1208.6477;%%
  %2 citations counted in INSPIRE as of 12 Mar 2013


\bibitem{KrY} D.\ Kreimer, K.\ Yeats,
 {\em Recursion and growth estimates in renormalizable quantum field theory,}
  Commun.\ Math.\ Phys.\  {\bf 279} (2008) 401
  [hep-th/0612179].
  %%CITATION = HEP-TH/0612179;%%

\bibitem{KrY6p} D.\ Kreimer, K.\ Yeats,
  {\em An Etude in non-linear Dyson-Schwinger Equations,}
  Nucl.\ Phys.\ Proc.\ Suppl.\  {\bf 160} (2006) 116
  [hep-th/0605096].
  %%CITATION = HEP-TH/0605096;%%

\bibitem{Cor}
%\cite{Kreimer:2012eh}
%\bibitem{Kreimer:2012eh}
  D.~Kreimer and K.~Yeats,
  {\em Properties of the corolla polynomial of a 3-regular graph,}
  arXiv:1207.5460 [math.CO], Electronic J.\ of Combinatorics {\bf 20(1)} (2013), \#P41.
  %%CITATION = ARXIV:1207.5460;%%
  %1 citations counted in INSPIRE as of 12 Mar 2013

\bibitem{PanMas}
%\cite{Panzer:2012gp}
%\bibitem{Panzer:2012gp}
  E.~Panzer,
  {\em Hopf algebraic Renormalization of Kreimer's toy model,}
  arXiv:1202.3552 [math.QA], Master Thesis at Humboldt University.
  %%CITATION = ARXIV:1202.3552;%%
  %1 citations counted in INSPIRE as of 12 Mar 2013

\bibitem{PanMadrid}
E.\ Panzer, {\em Renormalization, Hopf algebras and Mellin transforms }, 
{\tt http://www.mathematik.hu-berlin.de/~maphy/panzer.pdf}, these proceedings.


\bibitem{PanzerBingen} E.\ Panzer, talk at Spring School: Feynman Graphs and Motives, Bingen, March 2013,
{\tt http://www.sfb45.de/events/spring-school-feynman-graphs-and-motives}.

\bibitem{PanzerInPrep} 
  E.~Panzer,
  {\em On the analytic computation of massless propagators in dimensional regularization,}
  Nuclear Physics, Section {\bf B 874} (2013), pp. 567-593,
  arXiv:1305.2161 [hep-th].





\bibitem{GraphFun} O.\ Schnetz, {\em Graphical functions and single-valued multiple polylogarithms}, arXiv:1302.6445.




\bibitem{Ytheses} K.\ Yeats,
{\em Rearranging Dyson-Schwinger equations,} Memoirs of the AMS 211, arXiv:0810.2249 [math-ph].

\bibitem{YeatsMadrid}
%\cite{Yeats:2013au}
%\bibitem{Yeats:2013au}
  K.~Yeats,
  {\em Some combinatorial interpretations in perturbative quantum field theory,}
  arXiv:1302.0080 [math-ph], these proceedings.
  %%CITATION = ARXIV:1302.0080;%%


\end{thebibliography}

\end{document}